# An Energy Efficient Self-healing Mechanism for Long Life Wireless Sensor Networks

Dame Diongue
PhD Student, Department of Computer Science
Gaston Berger University
Saint Louis, SENEGAL
ddiongue.ep2112812@ugb.edu.sn

Ousmane Thiare
Department of Computer Science
Gaston Berger University
Saint Louis, SENEGAL
ousmane.thiare@ugb.edu.sn

*Abstract*—In this paper, we provide an energy efficient self-healing mechanism for Wireless Sensor Networks. The proposed solution is based on our probabilistic sentinel scheme. To reduce energy consumption while maintaining good connectivity between sentinel nodes, we compose our solution on two main concepts, node adaptation and link adaptation. The first algorithm uses node adaptation technique and permits to distributively schedule nodes activities and select a minimum subset of active nodes (sentry) to monitor the interest region. And secondly, we in- troduce a link control algorithm to ensure better connectiv- ity between sentinel nodes while avoiding outliers appearance. Without increasing control messages overhead, performances evaluations show that our solution is scalable with a steady energy consumption. Simulations carried out also show that the proposed mechanism ensures good connectivity between sentry nodes while considerably reducing the total energy spent.

*Keywords-component—Wireless Sensor Networks, Self-healing, Energy efficiency, Node adaptation, Link adaptation.*

## I. INTRODUCTION

Recent technological advances in microelectronics have favored the development of tiny and intelligent embedded devices called sensor nodes that can detect and send relevant informations related to a given environment. This has led to the emergence of a new technology, Wireless Sensor Networks. A typical Wireless Sensor Networks consists of a huge number of tiny sensor with sensing, processing and transmission capabilities [1]. These last decades, the wireless sensor technology holds the lead of the stage in several sec- tors such as environmental monitoring, military surveillance, medical diagnosis, building automation, industrial automation tasks, etc. In most cases, the area of interest (wireless sensor network's deployment area) is harsh or even impossible to access for human intervention. Therefore, the deployment is most often done by air plane dropping and this may often lead to unfair repartition of sensor nodes through the monitored region.

Beside problems related to random deployment, Wireless Sensor Networks are also suffering to many challenges such as data aggregation, routing, security, energy management, topology management, etc. The two later issues have attracted more and more interest from researchers and are addressed in this paper. Energy consumption and topology changes are of critical importance regarding Wireless Sensor Networks because the sensor node lifetime is closely related to its battery power and once deployed, they are usually inaccessible to be replaced nor recharged, due to harsh environment. However, the protocol designers should take into consideration these constraints and allow sensor nodes to have autonomy to self organize and covertly save their energy. In some types of applications, random deployment is most often used and it does not always guarantee better coverage and rational use of energy. This type of deployment, often causes energy or coverage hole problems due to unfair repartition of sensor nodes.

In this paper, we propose an energy efficient self-healing mechanism for wireless sensor networks. Our proposed mechanism use a combination of node adaptation and link adaptation techniques. The node adaptation technique consists of scheduling redundant nodes to take over sentinel nodes when they fall down. And the other one, link adaptation, is applied for connectivity check between sentinel nodes (active nodes).

The remainder of this paper is organized as follow. Section II, presents some related works in the literature. Section III, details our proposed mechanism. Section IV talks about simulation and experimental results. And finally, section V, presents our conclusion and future interested issues.

## II. RELATED WORKS

A Wireless Sensor Network well-functioning strongly de- pends on:

• A good coverage of the interest area to retrieve relevant information

• A good connectivity between sensor nodes to better relay information toward the Sink node

• And also a good energy management policy for a long life network

However, the deployment strategies (deterministic or random) have a great influence on above criteria. Ideally, a deterministic deployment is desirable, but in most cases the monitored region, for example battle field, is inaccessible or dangerous for human access. Thus, a random deployment remains the only alternative way to monitor such regions. This deployment method often leads to collateral problems such as sparse or not at all covered areas. During the last decade, several works have been done in topology control issues. Solutions has been proposed in the literature in order to solve the related problems to the network topology changes. And these solutions can be classified according three approaches: node adaptation, link adaptation and mobility (mobile sensor node or robot) [2]. Node adaptation techniques are often based on clustering which propose the network to have an hierarchical organization, set cover computation which organize the network into multiple subset where each one can cover the whole network for a period of time and lastly node scheduling technics that relies on deploying redundant nodes and schedule their activities.

In [3], inspired by Ant Colony Optimization, authors propose an algorithm for set cover computation and selection. They propose finding maximum disjoint connected cover sets that satisfy network coverage and connectivity between nodes within a set cover. The ant pheromone is used as a metaphor for a search experience in cover sets reconstruction. Always in the same direction, Gupta et al. [4] use a node scheduling technique for topology healing and a probabilistic approach to determine the coverage redundancy degree. They schedule nodes activities on the one hand to save energy and also ensure a better coverage. In [5], Corke et al. propose two algorithms. The first algorithm uses neighbor informations to detect failed nodes and determine hole location. The second algorithm uses routing informations to detect a hole from a distance and try to maintain the routing path. Their solution require that nodes keep state informations in memory.

Another approach in the literature is link adaptation based technic. Link adaptation is one of the common techniques used for topology control in Wireless Sensor Networks by adapting communication parameters and exchanging neighbor informations. As shown in [6], authors propose a self- healing framework based on link quality measurement to detect and solve coverage holes. They use dependency constraints through a three modules framework (Health Monitoring module, Self-healing Policy and Self-healing Engine). Li et al. [6] tune the received signal strength in order to maintain communication path and in the same way save energy when possible. In [7], the authors follow the same vein and propose a self-stabilizing algorithm for efficient topology control. They perform nodes' communication range adjustment by reducing when necessary the transmission power an build connected dominating set in order to establish efficient routing path. Cerulli et al. propose in [8] an appropriate set covers activation time scheduling. At every set cover construction, they make a procedure that select adjusted sensors and avoid banned other. They define banned sensors and exclude them in the set cover selection to avoid target lost.

Wang et al. [9] surveyed the mobility approach for topology healing in wireless sensor networks. Two main mobility strategies can be listed: mobility of sensor nodes [10]–[12] and the deployment of additional mobile robots [13]. Ghosh et al. show in [10] that Voronoi Diagrams can be used to detect and heal topology change problems like coverage or energy holes. Using the nodes' overlapping sensing range, authors propose a deterministic deployment of additional mobile nodes to maintain the network (coverage and connectivity maintenance). Authors in [12] follow in the same vein by providing additional mobile nodes to cover the holes. They define different roles assigned to the nodes and that role definition is based on node's residual energy and its location relatively to the hole. Then they propose an election method to guide the choice of the leader node which have the responsibility to interact with mobile nodes and designate when and where they should be deployed. Their solution does not take into account a sudden node failure in particular the leader node failure. If the leader node suddenly crash, the whole recovery process will be compromised. Bo-Ruei et al. [11] on their part, propose a dynamic multihop topology healing algorithm based on node mobility. Their algorithm is composed of two phases per round: construction and moving phases. The construction phase involves collecting neighborhood informations. This allows the nodes to have knowledge of the network topology from which they determine the target points. And then comes the moving phase which guide with precision (direction and distance) the movement of mobile nodes. This solution generates a lot of overhead with the exchanged neighbor informations. Wu et al. [14] make a deep analysis of energy hole problems and propose a distribute coronas based deployment to avoid energy holes.

In this paper, we opted for a scheduling based solution rather than deploying additional mobile nodes. Because energy is a scarce resource in Wireless Sensor Networks. However, mobility based solutions, in addition to the expensive costs of equipments, use GPS (Global Position System) and mobilizer components which are very energy intensive. And also, mobility is often not easy or not at all applicable to some regions because of their relief.

### III. PROPOSED MECHANISM

#### A. Problem Description

A sensor network should be autonomous for topology healing for two key reasons:

- The area of interest is often harshly or simply inaccessible.

- The lifetime of a node strongly depends on its battery which has a limit power.

Hence, Wireless Sensor Network protocol designers need to take into consideration mechanisms that would make autonomous networks in term of topology changes. In order to fulfill network surveillance functionalities, the sensor nodes must be organized to cover the whole monitored region. And the period of time that the network still satisfy this condition is generally defined as the network's lifetime. As a fundamental research issue, topology control is attracting more and more researchers. Wireless Sensor Networks are faced to several challenges such as coverage holes often caused by topology change or deployment. Topology changes in Wireless Sensor Networks may be caused either by energy depletion or by a key module failure (sensor unit, transmission unit, etc.). Without a coverage healing procedure, nodes failure can cause partial or total path loss toward the sink node or simply coverage holes (see example in Fig. 1). In this scenario, all the sentry are connected at time period $t$ (*Fig. 1. (a)*) and after a while, i.e at time period $t+\Delta t$ sentry nodes *S4* and *S2* died, leaving their dedicated area blank (*Fig. 1. (b)*). To solve this problem, protocol designer must take into account these eventual changes, by proposing distributed and dynamic energy efficient algorithm fitted for WSNs, to overcome these challenges. On our part, we propose an energy efficient self-healing mechanism for wireless sensor networks. Our solution combines two techniques, at different levels, (link and node adaptation) among those mentioned above. The link adaptation technic is applied to active nodes (sentry), this is to avoid coverage holes appearance between sentinel nodes, while node adaptation technic is applied to reserve nodes (redundant nodes subject to pass occasionally into probing mode).

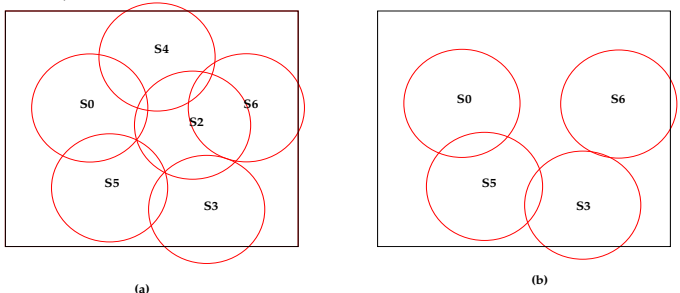

Fig. 1. Coverage loss and holes appearance

### B. Energy Efficient Design

In this paper, we propose an energy efficient topology healing mechanism based on a probabilistic model and nodes redundancy. We choose to deploy in the area of interest a huge number of sensor nodes such a way to create a high redundancy. And each sensor node had sufficient autonomy to compute and control it sleep and wake up phases. At the deployment, all sensor nodes are initially in sleep mode and that for a Weibull distribution time [15]. When they wake up, nodes will compete with the neighborhood to choose a sentinel which will be responsible for the monitoring. When a node wakes up, it probe its neighborhood to look for a standing sentinel. If it is aware that there is no guard present in the vicinity, it immediately stands guard (gets a sentinel role) and monitors the dedicated area. Else, it updates its probe rate for a new sleep time computation and then goes back to sleep mode.

### C. Hole healing through link adaptation

Due to the nature of the carrier, collisions may disrupt wireless sensor nodes' communications. Therefore, nodes can wake up and start guard, after probing, believing that there is no sentinel node in its vicinity. This may results to a duplicated monitoring in a given region. To avoid this energy waste, we have chosen in this work to combine node adaptation technic [15] with link adaptation one. Here, the link adaptation technique allow us to dynamically adjust a sentinel node's communication parameters. As shown in *Algorithm 1*, sentinel nodes use connectivity message to evaluate the link quality between them. Thus, each sentinel node randomly shot a timer and when it expires, it sends connectivity message to other sentinel nodes. And so, when they receive the message, they send back (by unicast) a reply. The sentinel node which initiated the dialogue, measure the link quality based on the obtained LQI[1] [16,17]. If the measured LQI value is below to a given threshold, node will adjust its communication power strength in the aim to maintain a good connectivity with other sentinel nodes. Otherwise, it ignores the message and prepares itself for another connectivity check round.

```
Algorithm 1 Holes healing using active nodes (Connectivity
adjustment between sentinel nodes)
1:  t_c ≠ 0, linkState = {strong, weak}
2:  status = ACTIVE, connMsg = FALSE
3:  if (t_c expires) then
4:      Send connectivity msg
5:      if (connMsg) then
6:          Check linkrobustness(LQI) from received msg
7:          if (linkState == strong) then
8:              Set timer t_c
9:          else
10:             Adjusts link parameters
11:             Set timer t_c
12:         end if
13:     end if
14: end if
```

### D. Hole healing through node adaptation

In this paper we propose a dynamic and distributed topology healing mechanism. Our approach enable nodes to have sufficient autonomy to control the network topology by having a dynamic and distributed holes detection. Nodes have complete control over when they go to sleep or not. For this, we consider a sufficiently dense network and we propose to exploit nodes redundancy on the one hand extend network lifetime and on the other hand compensate coverage holes often due to node failures. Our hole healing scheme operate at two levels: coverage recovery by reserve nodes and connectivity adjustment by sentinel nodes. The reserve nodes (redundant nodes) sleep most over the time to save their energy. But, they often wake up, according to a Weibull sleep time computation, to probe their neighborhood (refer to the

---
[1] Link Quality Indication - CC2420

*Algorithm 2*.). The probing phase permits, when the neighboring sentinel falls down, that a reserve node detect this failure and wakes up to take it over for the monitoring task. The scenario in *Fig. 2* is an illustration example. At the period $t$, the subset of sentinel nodes is composed by $S_0$, $S_1$, $S_8$, $S_{13}$, $S_{15}$ and $S_{20}$. After a while, nodes $S_0$ and $S_8$ fall down and then, the reserve nodes $S_2$ and $S_7$, during their probing phase detect the problem and wake up to maintain the topology. Thus, at time period $t + \Delta t$, we have $S_1$, $S_2$, $S_7$, $S_{13}$, $S_{15}$ and $S_{20}$ in guard as sentinel nodes.

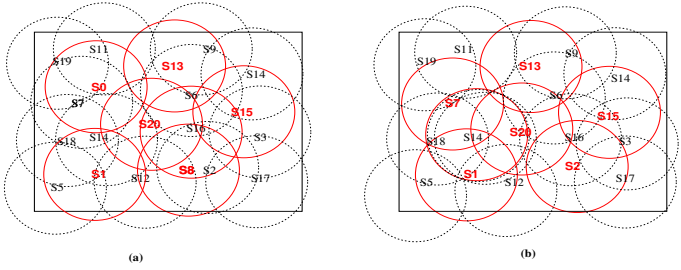

Fig. 2. Hole healing with redundant nodes

**Algorithm 2** Hole maintenance using redundant nodes (nodes switching between sleep and probing mode looking for on guard sentinel neighbor). status is a node's mode; $t_s$, sleep time; $t_w$, wait for neighbor response time; $\lambda$, Weibull scale parameter; $\beta$, Weibull shape parameter

1: $status = SLEEP, rcvMsg = FALSE$
2: $t_s > 0, \lambda > 0, t_w > 0$
3: **if** ($t_s$ expires) **then**
4:    $status = PROBE$
5:    probe neighborhood
6:    set timer $t_w$
7:    **if** ($t_w$ expires) **then**
8:      **if** ($rcvMsg$) **then**
9:        $R = uniform(0, 1)$
10:       $t_s = \frac{1}{\lambda} \log\left(\frac{1}{R}\right)^{\frac{1}{\beta}}$
11:       $\lambda(t) = (\beta \times \lambda) \times (t \times \lambda)^{\beta-1}$
12:       $status = SLEEP$
13:       set timer $t_s$
14:      **else**
15:        $status = ACTIVE$
16:        generate random timer for
17:        connectivity check $t_c$
18:      **end if**
19:    **end if**
20: **end if**
21:

## IV. SIMULATIONS AND RESULTS

To evaluate our proposed solution's performance, simulations are carried out using Castalia [18], an OMNeT++ [19] framework dedicated to Wireless Sensor Networks (WSNs) and Body Area Networks (BANs). We checked expected improvement of our solution by comparing it to our previous works in [15].

### A. Simulation setup

Simulations are performed in an interest area of 100×100 m² with a number randomly deployed nodes from 50 to 1000. The simulation parameters details are summarized in Table I.

TABLE I
SIMULATION PARAMETERS

| Field | 100 x 100 meters |
|---|---|
| Number of nodes | [50:1000] |
| Deployment type | uniform |
| Environment | Noisy |
| Nodes' transmission power | -5 dBm, -10 dBm |
| LQI threshold | 7 |

### B. Results and discussions

This work focus on self-healing solution for Wireless Sensor Networks by emphasizing the energy efficiency and strong connectivity between sentry.

1) *Reliable sentry connectivity:* To maintain strong connectivity between sentinel nodes, the *Algorithm 1* performs link quality detection in aim to avoid sentinel outlier appearance. We include link quality check when exchanging probe messages. That is say, if a sentinel node receives a probe response message, it check the LQI value and compares it with a desired one. When it detects a weak link *i.e.* LQI test fails, it adjusts its transmission power from −10 dBm to −5 dBm. And *Fig. 3* can illustrate the obtained connectivity between sentry nodes.

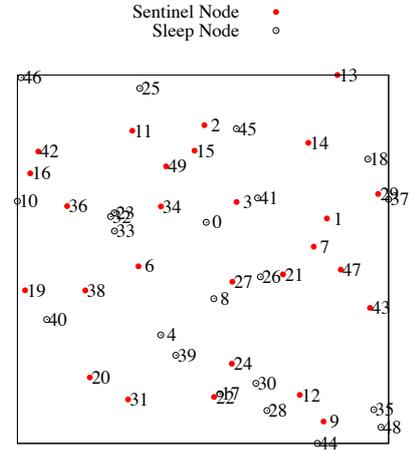

Fig. 3. Hot network snapshot with 50 rabdomly deployed nodes

2) *Energy efficiency:* The average energy spent is measured under two scenario:

- Average energy consumption by varying the Weibull shape parameter β. We recall that our sentinel scheme uses the Weibull probabilistic distribution to calculate reserved nodes wake up times and also to update their probe rates. Curves in *Fig. 4* show that, with the enhancement introduced here,the energy consumption is not affected when varying the β parameter. This shows that our mechanism provides an almost steady energy consumption compared to other probability distributions, variant of the Weibull probability distribution, such as Rayleigh when β = 2.

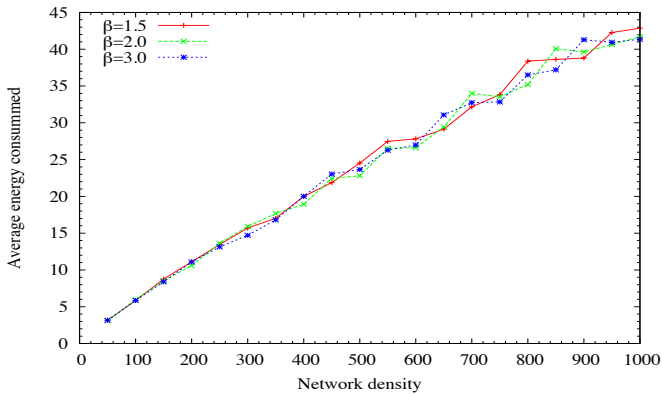

*Fig. 4. Average energy consumption by varying the β parameter*

- Curves in *Fig. 5* show total energy consumption of the whole network after 1000 seconds of simulation time. We note that our algorithm, upon maintaining sentry connectivity, considerably reduce the average energy consumption. This can be justified by the fact that we opted to include the link control step in the probing downward phase. This is to say that, the link control algorithm is called when sentry node receive probe response from others sentry. This permits by reducing control overhead (reduce in the same time transmissions between sentry nodes) to considerably improve energy consumption.

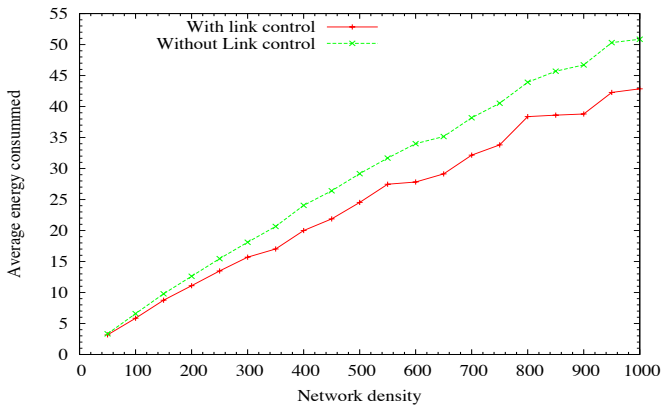

*Fig. 5. Average energy consumption for sentinel scheme: link control vs. no link control*

Another advantage of our solution is that it supports the network scalability without a remarkable impact in the total consumed energy.

V. CONCLUSION

In this paper, we focus on topology healing solutions through distribute sleep scheduling. We propose an energy efficient self-healing mechanism designed to scheduling sensor nodes activities and also prune duplicated sentinel nodes for a single area monitoring. Performances evaluation show that our proposed solution performs strong connectivity between sentry node, permits a flexible and efficient energy usage and finally it supports the network scalability.

Future works will targeted the robustness of our scheme against security attacks like DoS (Deny of Service) or Deny of Sleep attack in Wireless Sensor Networks.